\documentclass[a4paper,12pt]{article}

\usepackage[english]{babel}
\usepackage[utf8x]{inputenc}
\usepackage[T1]{fontenc}

\usepackage[a4paper,top=3cm,bottom=2cm,left=3cm,right=3cm,marginparwidth=1.75cm]{geometry}

\usepackage{amsmath}
\usepackage{graphicx}
\usepackage[colorinlistoftodos]{todonotes}
\usepackage[colorlinks=true, allcolors=blue]{hyperref}

\usepackage{graphicx}
\usepackage{subcaption}
\usepackage{float}

\title{Scaling of Spoke Rotation Frequency within a Penning Discharge}

\author{
  Andrew T. Powis\\
  \textit{Princeton University, Princeton, New Jersey 08544, USA}\\
  Johan A. Carlsson, Igor D. Kaganovich, Yevgeny Raitses\\
  \textit{Princeton Plasma Physics Laboratory, Princeton, New Jersey 08540, USA}\\
  Andrei Smolyakov\\
  \textit{University of Saskatchewan, Saskatoon, Saskatchewan S7N 5E2, Canada}
}

\date{}

\begin{document}
\maketitle

\begin{abstract}
A rotating plasma spoke is shown to develop in two-dimensional full-sized kinetic simulations of a Penning discharge cross-section. Electron cross-field transport within the discharge is highly anomalous and correlates strongly with the spoke phase. Similarity between collisional and collisionless simulations demonstrates that ionization is not necessary for spoke formation. Parameter scans with discharge current $I_d$, applied magnetic field strength $B$ and ion mass $m_i$ show that spoke frequency scales with $\sqrt{e E_r L_n / m_i}$, where $E_r$ is the radial electric field, $L_n$ is the gradient length scale and $e$ is the fundamental charge. This scaling suggests that the spoke may develop as a non-linear phase of the collisionless Simon-Hoh instability.
\end{abstract}

\section{Introduction}
\label{introduction}

Hall plasmas, consisting of magnetized electrons and unmagnetized ions, exhibit a wide range of plasma instabilities \cite{choueiri2001plasma,chesta2001characterization}. In some systems these instabilities may result in the formation of a long wavelength, low frequency fluctuation in plasma density, propagating in the $\mathbf{E} \times \mathbf{B}$ direction. This rotating structure is commonly known as the plasma ``spoke''.

The spoke has been well characterized within a number of experimental devices, including Hall thrusters \cite{janes1966anomalous,esipchuk1974plasma,parker2010transition,ellison2011direct,mcdonald2011rotating,ellison2012cross,griswold2012feedback,raitses2012studies,liu2014ultrahigh,sekerak2015azimuthal,cappelli2016coherent,hyatt2017hall,mazouffre2017investigation}, planar magnetrons \cite{anders2012drifting,winter2013instabilities,hecimovic2015spoke,poolcharuansin2015use,anders2017direct,hecimovic2017sputtering,panjan2017plasma,hnilica2018effect}, cylindrical magnetrons \cite{ehiasarian2012high,brenning2013spokes} and Penning discharges \cite{thomassen1966turbulent,sakawa1993excitation,raitses2015effects}. These devices feature cylindrical geometry, with electron drift and spoke propagation occurring in the azimuthal direction.

The plasma spoke may play an important role in the anomalous transport of electrons across the applied magnetic field, a source of reduced efficiency \cite{janes1962electrostatic}. Using separated probes within a Hall thruster, Janes \& Lowder \cite{janes1966anomalous} measured an azimuthal electric field correlated with the passage of the spoke and suggested that the resulting $\mathbf{E}_{\theta} \times \mathbf{B}_{r}$ drift was enhancing electron transport towards the thruster anode. More recently, the use of segmented anodes within a Cylindrical Hall Thruster \cite{raitses2001parametric} demonstrated that half of the discharge current was being carried through the spoke, evidence for enhanced transport through the structure \cite{ellison2011direct,ellison2012cross}.

Due to its highly non-linear, turbulent and global nature, the spoke has continued to evade a clear theoretical understanding. Proposed mechanisms include a type of ionization wave \cite{janes1966anomalous}, whereby the azimuthal field at the front of the spoke provides sufficient energy to ionize neutrals and propagate the plasma density perturbation. The ionization wave would propagate at the Critical Ionization Velocity (CIV) \cite{alfven1967origin} defined as $v_{civ}=\sqrt{2 U_{ion}/m_n}$, where $U_{ion}$  and $m_n$ are the ionization energy and mass of the neutral species respectively.

Alternatively, the spoke may be the result of collective effects, with the Collisionless Simon-Hoh Instability (CSHI) being a likely candidate \cite{thomassen1966turbulent,sakawa1993excitation}. Charge separation between drifting electrons and unmagnetized (non-drifting) ions generates an azimuthal electric field. If the background electric field and density gradients within the system are aligned $\mathbf{E}_{0} \cdot \nabla n_0 > 0$, then the the resulting azimuthal field will act to enhance perturbations in plasma density, driving the instability.

Investigation of the plasma spoke within a Penning discharge offers several advantages, most immediately being improved access for diagnostics. The applied field is ideally uniform and aligned with the axis of the device, allowing the electron dynamics parallel to the device axis to be decoupled from those in the transverse direction. For purposes of simulations and theory the system can therefore be treated two-dimensionally. The lack of a magnetic field gradient reduces the number of energy sources for instabilities, eliminating possible driving mechanisms for spoke formation. In many systems the plasma is also weakly collisional, further simplifying theoretical considerations. This makes it an ideal system within which to study spoke formation and anomalous transport.

Simulations have played an increasingly important role in understanding the formation of the spoke and its connection to anomalous transport. Kinetic techniques are required to self-consistently capture transport effects, with the Particle-in-Cell Monte-Carlo Collision method (PIC-MCC) commonly being used. The challenge with PIC-MCC comes from the numerical requirement to resolve the smallest time and length scales associated with the plasma, the electron plasma frequency and Debye length respectively. These constraints result in a significant computational overhead required to capture the vastly multi-scale physics observed within devices such as the Penning discharge.

This challenge is usually dealt with in one of two ways. The first approach is to scale the system in some way, either by reducing device size (to reduce the grid size), decreasing the plasma density or increasing the relative permittivity (to increase the cell size). With improving computational capabilities and employing scaling it has finally become possible to simulate a 2D profiles of a device \cite{boeuf2013rotating,boeuf2014rotating,escobar2014analysing,carlsson2015multi,escobar2015global,escobar2015low,carlsson2017particle,carlsson2018particle}, and scaled 3D devices \cite{matyash2012numerical,taccogna2011three,taccogna2012physics,matyash2013particle}. Low mode number, rotating structures have been observed in a number of these simulations, in particular, Matyash found close agreement between the frequency of the rotating structure observed within a model of the wall-less Hall thruster \cite{matyash2017investigation} experiment at CRNS \cite{mazouffre2017investigation}.

The second technique adopts a hybrid approach, modeling electrons as a fluid, such that the associated length and time scales need not be resolved. The primary disadvantage of this technique is that kinetic electron transport effects are not captured self-consistently and must be incorporated via a model informed from experimental evidence. None-the-less hybrid models have had success in reproducing the spoke. Most recently a two-dimensional axial-azimuthal hybrid code captured the motion of the spoke within a simulated thruster channel with SPT-100 like parameters. \cite{kawashima2018numerical}. The spoke velocity was on the order of what has previously been observed in experiments and suggested to be the result an instability driven by magnetic field and density gradients \cite{esipchuk1974plasma,frias2013long}.

Despite these capabilities, significant work is required to produce unscaled, fully kinetic simulations of entire devices. Perhaps equally as challenging will be developing tools for analysing the enormous amount of data produced by such simulations. 

The present paper is a follow up of the work commenced in Reference \cite{carlsson2018particle} whereby large scale coherent structures were observed within two-dimensional PIC-MCC simulations of a Penning discharge cross-section. These simulations are modified to allow quasi-steady-state operation of the discharge revealing formation of a single mode, azimuthally rotating structure. By comparing this structure with that observed within experiments, and studying how it scales with discharge properties, an attempt is made to determine the fundamental mechanism responsible for its formation.

\section{Methodology}

Simulations were run using the Large-Scale Plasma code (LSP) \cite{hughes1999three}. LSP is a multi-purpose and versatile PIC code, widely benchmarked and validated within the community \cite{hughes1999three,welch2001simulation,welch2006integrated,welch2009fully}. Since the plasma is sufficiently low temperature ($T_e \sim 5$ eV), the system can be treated electrostatically, and Poisson's equation inverted to obtain the electric potential from charge density. PPPL modifications to the code \cite{carlsson2016validation} include incorporation of the latest version of the Portable Extensible Toolkit for Scientific Computation (PETSc) \cite{petsc-web-page,petsc-user-ref,petsc-efficient} for improved performance and scalability. Poisson's equation was inverted via PETSc's native LU factorization package.

\begin{figure}[H]
\centering
\includegraphics[width=0.9\linewidth]{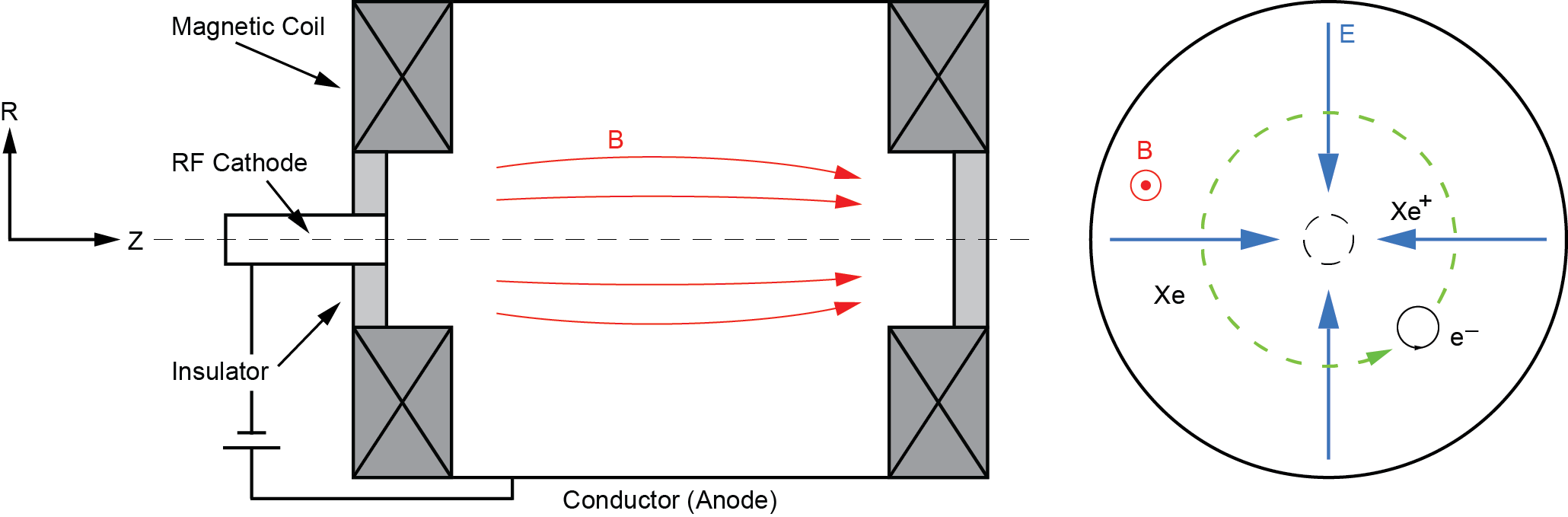}
\caption{\label{fig:penning_setup}a) Experimental setup of the Penning discharge at PPPL \cite{raitses2015effects} in $r$-$z$ geometry. Magnetic field lines are indicated in red. b) Cross-section of the Penning discharge, representative of the simulation domain in $r$-$\theta$ geometry. Blue lines indicate the radial ambipolar electric field and green-dashed lines represent the direction of the electron $\mathbf{E} \times \mathbf{B}$ drift. Xenon ions are unmagnetized.}
\end{figure}

Simulation were modeled off the Penning discharge experiment at the PPPL-HTX laboratory \cite{raitses2015effects}. Within this device an RF plasma cathode injects electrons along the axial magnetic field, ionizing a low pressure gas of either Argon or Xenon. Electron motion in the radial direction is inhibited by an axially applied uniform magnetic field. Ions are weakly magnetized and therefore significantly more mobile, giving rise to an ambipolar radial electric field. The applied magnetic field and ambipolar electric field result in electrons undergoing $\mathbf{E}_{r} \times \mathbf{B}_{z}$ drifts in the azimuthal direction. The plasma is surrounded by a grounded cylindrical metal anode with $10$ $cm$ diameter. The ends of the cylinder are dielectric, preventing the short-circuit effect. The device geometry and simulation geometry are shown in Figure \ref{fig:penning_setup}.

A slice of the device cross-section was modeled on a uniform Cartesian grid, with particles evolving in 2D-3V phase space. A Cartesian grid was chosen over a cylindrical grid to avoid numerical instabilities associated with the grid singularity at zero radius. A uniform magnetic field is applied perpendicular to the simulation domain. To shed light on the formation mechanism of the spoke, both collisionless and collisional simulations were performed. In collisionless simulations, electrons and ions are injected into the center of the trap at fixed current and initial temperature. For collisional simulations, electrons are injected into a uniform background of neutrals, with ions forming via ionization. Modeled collisions include Coulomb collisions between all charged species and electron-neutral collisions. Neutral particle excitation was not modeled. Both electron-neutral elastic and ionizing collisions were informed by experimental data \cite{rapp1965total}.

Simulations were designed to be completed within a realistic time frame, enabling parametric investigations of instabilities within the Penning discharge. To achieve this goal the simulation domain was reduced from $10$ $cm$ to $5$ $cm$ and the Xenon gas mass was reduced to that of Helium-4. To relax the strict constraints on PIC simulations the relative permittivity was increased to $\varepsilon_{r}=400$, increasing the Debye length and reducing plasma frequency, thereby allowing for larger cell size and time step respectively. The relative permittivity was scaled, since it is suspected to play a small role in the instabilities which may be responsible for spoke formation. This led to a grid of $250 \times 250$ cells, with cell size $\Delta x = 200$ $\mu m$ and time step $\Delta t = 40$ $ps$, suitable to resolve all relevant plasma length and time scales. Simulations of $100$ $\mu s$ of plasma time could therefore be completed within 2 days. Simulation were run with 28 cores on the Princeton University Perseus supercomputer and 24 cores on the Department of Energy’s National Energy Research Scientific Computing Center (NERSC) Edison supercomputer. Post-processing was performed using the IDL based LSP post-processing tool P4 and in-house \textit{Python} codes.

\section{Results \& Discussion}

\subsection{Collisionless Simulations of the Rotating Spoke}
\label{first_spoke}

Electrons and ions are injected into the center of an initially empty domain with fixed currents and temperature $I_e$, $T_{e,inj}$ and $I_i$, $T_{i,inj}$ respectively (see Table \ref{tab:sim_params}).

\begin{table}[H]
\centering
\caption{\label{tab:sim_params}Physical parameters of simulation.}
\begin{tabular}{|c|c|c|c|}
\hline
Property & Symbol & Value & Units \\\hline
Relative Permittivity & $\varepsilon_r$ & $400$ & $-$ \\
Discharge Radius & $R_0$ & $2.5$ & $cm$\\
Applied Magnetic Field & $B_0$ & $100$ & $Gauss$\\
Electron Current & $I_e$ & $250$ & $mA$\\
Ion Current & $I_i$ & $100$ & $mA$\\
Discharge Current & $I_d$ & $-150$ & $mA$\\
Electron Injection Temperature & $T_{e,inj}$ & $5$ & $eV$\\
Ion Injection Temperature & $T_{i,inj}$ & $293$ & $K$\\
Electron Beam Energy & $V_{b}$ & $15$ & $eV$\\
Neutral Pressure & $P_n$ & $1$ & $m Torr$\\
Neutral Temperature & $T_{n}$ & $293$ & $K$\\
Electron-Neutral Cross-Section & $\sigma_{en}$ & $2.88 \times 10^{-19}$ & $m^2$\\\hline
\end{tabular}
\end{table}

The negative discharge current $I_d=I_i - I_e < 0$, results in an initially non-neutral plasma until sufficient ions have been injected to provide a neutralizing background. Quasi-neutrality is achieved around $100$ $\mu s$, after which a long-wavelength, single mode structure forms, rotating in the azimuthal direction. When viewed in the direction of the applied magnetic field, the structure rotates in the anti-clockwise direction. 

\begin{figure}[H]
\centering

\begin{subfigure}[b]{1.0\linewidth}
\includegraphics[width=\linewidth]{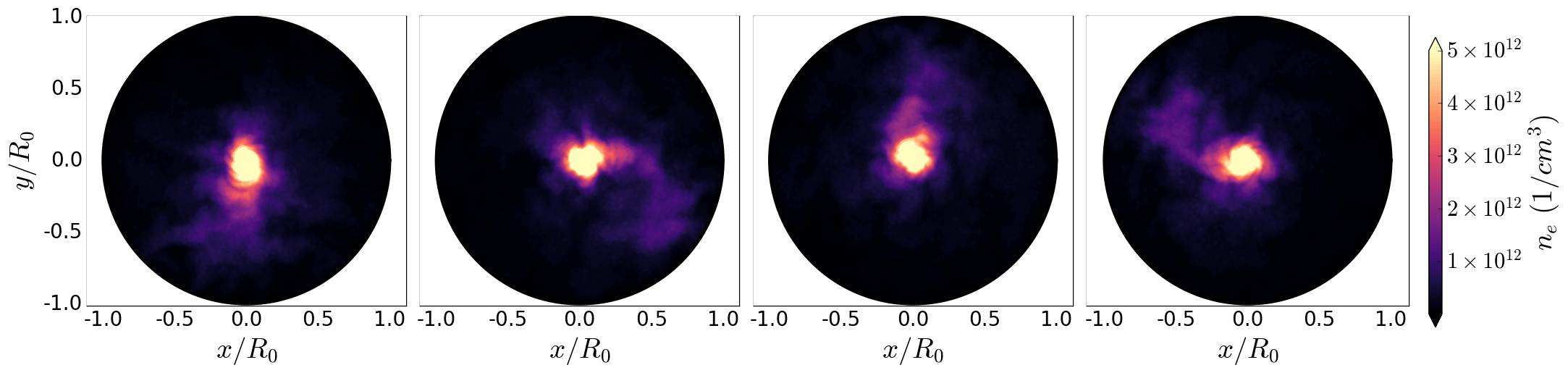}
\caption{}
\label{fig:lin_cless_a}
\end{subfigure}

\begin{subfigure}[b]{1.0\linewidth}
\includegraphics[width=\linewidth]{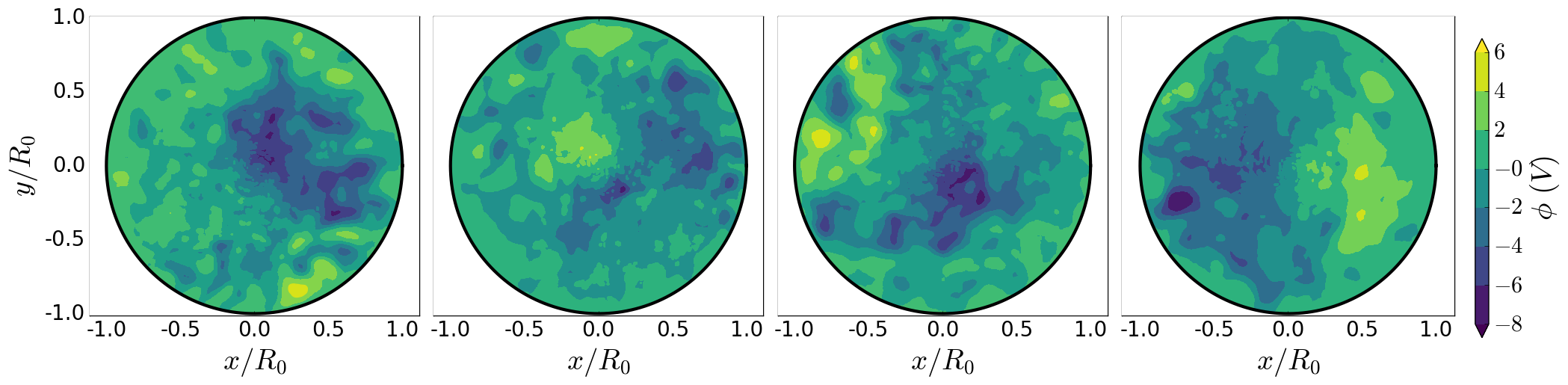}
\caption{}
\label{fig:lin_cless_b}
\end{subfigure}

\begin{subfigure}[b]{1.0\linewidth}
\includegraphics[width=\linewidth]{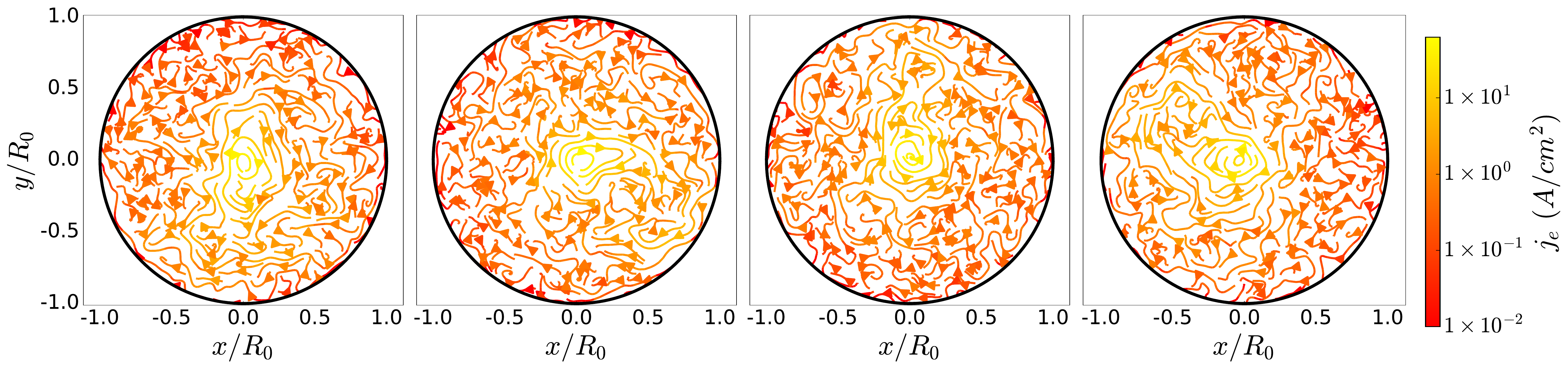}
\caption{}
\label{fig:lin_cless_c}
\end{subfigure}

\caption{\label{fig:lin_cless}a) Electron density contours, b) plasma potential contours, c) current streamlines, of the collisionless Penning discharge at simulation time, from left to right; $250$ $\mu s$, $253.8$ $\mu s$, $257.5$ $\mu s$ and $261.3$ $\mu s$.}
\end{figure}

Figure \ref{fig:lin_cless} shows contour plots of instantaneous electron density, plasma potential and current streamlines, commencing at $250$ $\mu s$ and incremented by a $\pi/4$ phase shift of the rotating structure. The structure does not appear to rotate as a rigid body, but rather as a density perturbation rich in micro-instabilities and turbulence. This is further evidenced by Figure \ref{fig:lin_cless_b} which shows a noisy and highly fluctuating plasma potential, with dips in plasma potential only weakly correlated with the rotating density perturbation. Figure's \ref{fig:lin_cless_b} and \ref{fig:lin_cless_c} show that there is neither a strong electric potential gradient nor current channel associated with the front of the spoke, however both indicate the presence of a turbulent plasma. Qualitatively it appears that increased cross-field electron transport within the structure is the result of enhanced turbulence, rather than a strongly correlated azimuthal electric field.

The phase of the rotating structure is measured to obtain a rotation frequency of $f_s = 66.0$ $kHz$, significantly lower than the electron and ion plasma frequencies, as well as the electron cyclotron frequency and the lower-hybrid frequency. When computed at $r=R_0/2$, the structure rotation velocity is $v_r = 5.18$ $km/s$, around half of the ion-acoustic velocity $v_s = 10.6$ $km/s$.

Numerical convergence was verified by measuring the mode frequency for simulations with half the cell size, half the time step and double the particle number, showing no more than a 3.8\% discrepancy. Modeling convergence was checked by scaling the relative permittivity, which for $\varepsilon_0 = 100$ agrees within 11 \% (see Table \ref{tab:verify}).

\begin{table}[H]
\centering
\caption{\label{tab:verify}Spoke frequency for numerical and modeling convergence studies.}
\begin{tabular}{|c|c|c|}
\hline
Modification & Spoke Frequency $(kHz)$ & Discrepancy $\%$\\\hline
$\Delta t/2$ & 68.5 & 3.8\\
$\Delta t/2$ \& $\Delta x/2$ & 63.9 & 3.2\\
Double Particle Number & 64.2 & 2.7\\
$\varepsilon_r = 100$ & 72.9 & 10.5\\\hline
\end{tabular}
\end{table}

The discharge remains in a quasi-steady state, with the rotating structure persisting until the end of the simulation at $500$ $\mu s$. This comprises approximately $26$ full rotations, from which time averaged statistics of the turbulent structure can be computed. Average radial profiles are computed by first azimuthally averaging each sample, and then temporally averaging over all samples. To compare the effects of the relative permittivity $\varepsilon_r$ on the plasma properties, averaging is performed for values of $\varepsilon_r =\{100,400\} $.

\begin{figure}[H]
\centering
\begin{subfigure}[b]{.45\linewidth}
\includegraphics[width=\linewidth]{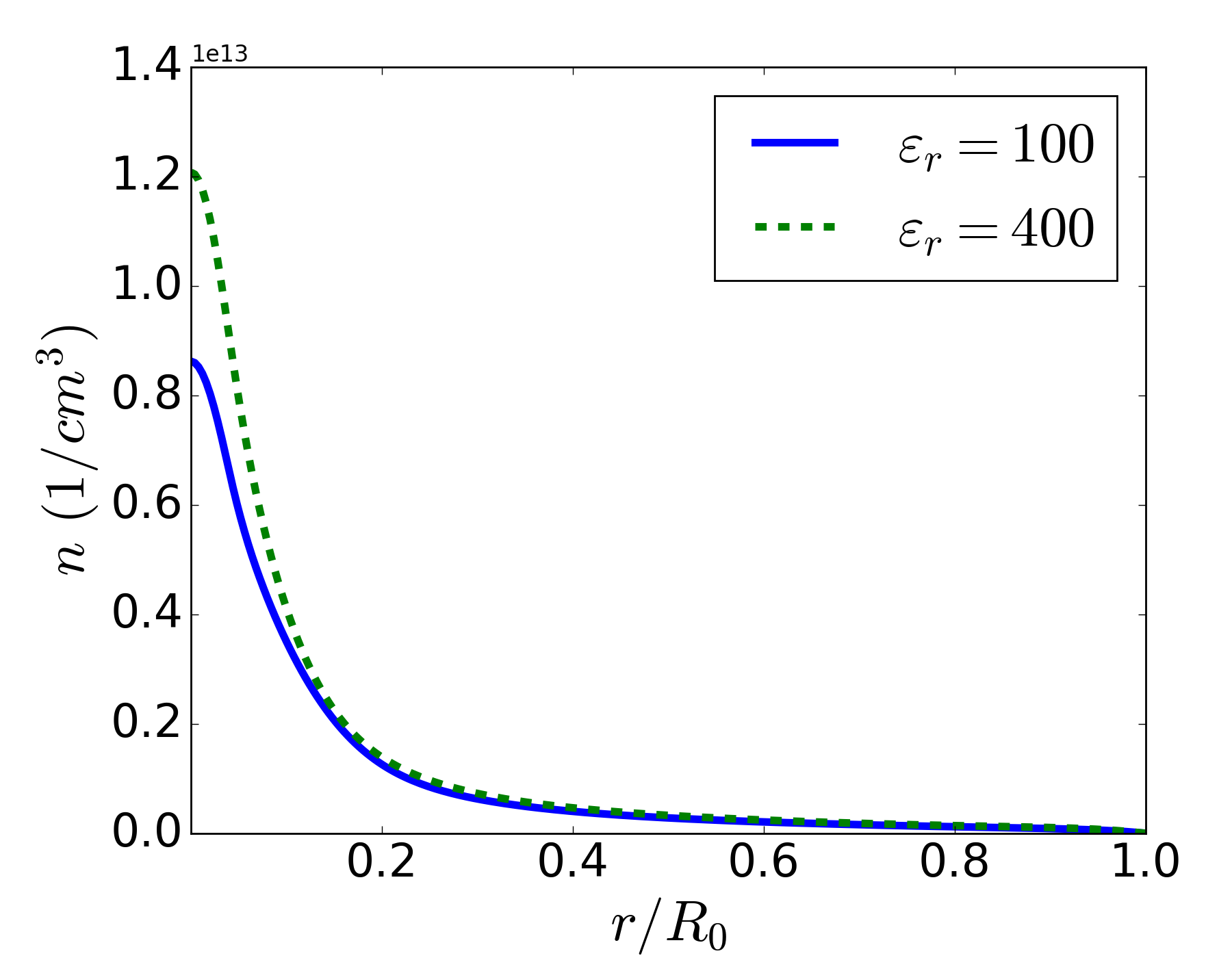}
\caption{\label{fig:avg_density}}
\end{subfigure}

\begin{subfigure}[b]{.45\linewidth}
\includegraphics[width=\linewidth]{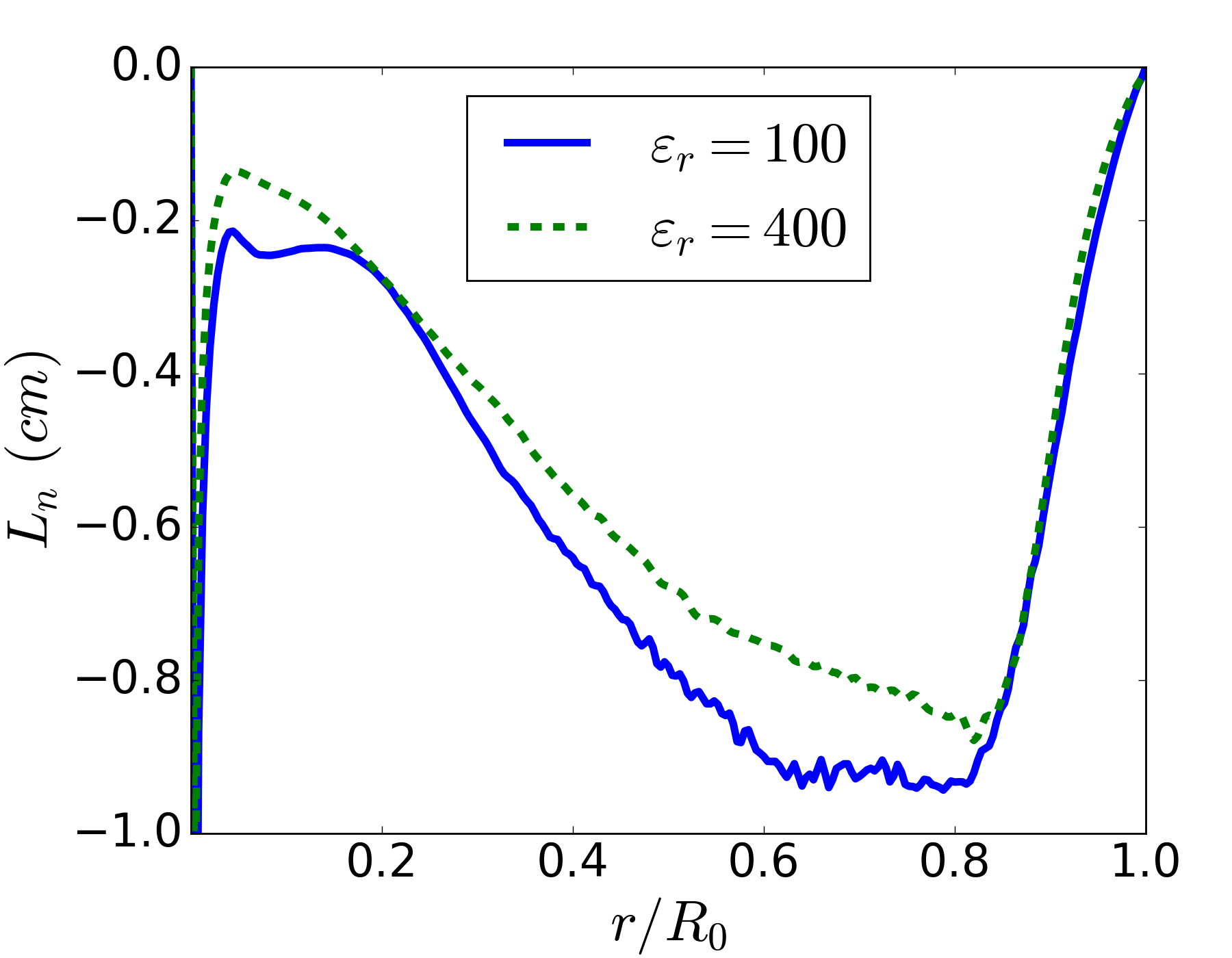}
\caption{\label{fig:avg_gls}}
\end{subfigure}

\caption{Azimuthally and temporally averaged radial profiles of a) electron density $n_e(r)$, b) gradient-length-scale $L_n(r) = n_e(r)/(dn_e(r)/dr)$ for $\varepsilon_r = \{100,400\}$.}
\label{fig:avg_density_plots}
\end{figure}

Figures \ref{fig:avg_density} and \ref{fig:avg_gls} show the averaged radial profiles of electron density $n_e(r)$ and gradient length scale $L_n(r) = n_e(r) / (dn_e(r) / dr)$. The plasma density is peaked at the center of the trap due to the source of electrons and ions, decaying away radially as the electrons are transported across the applied axial magnetic field. Increasing the relative permittivity results in a heightened density peak at the center of the trap, possibly due to weaker fields and a reduced rate of cross-field transport. The gradient-length-scale is negative at all radial positions.

\begin{figure}[H]
\centering
\begin{subfigure}[b]{.45\linewidth}
\includegraphics[width=\linewidth]{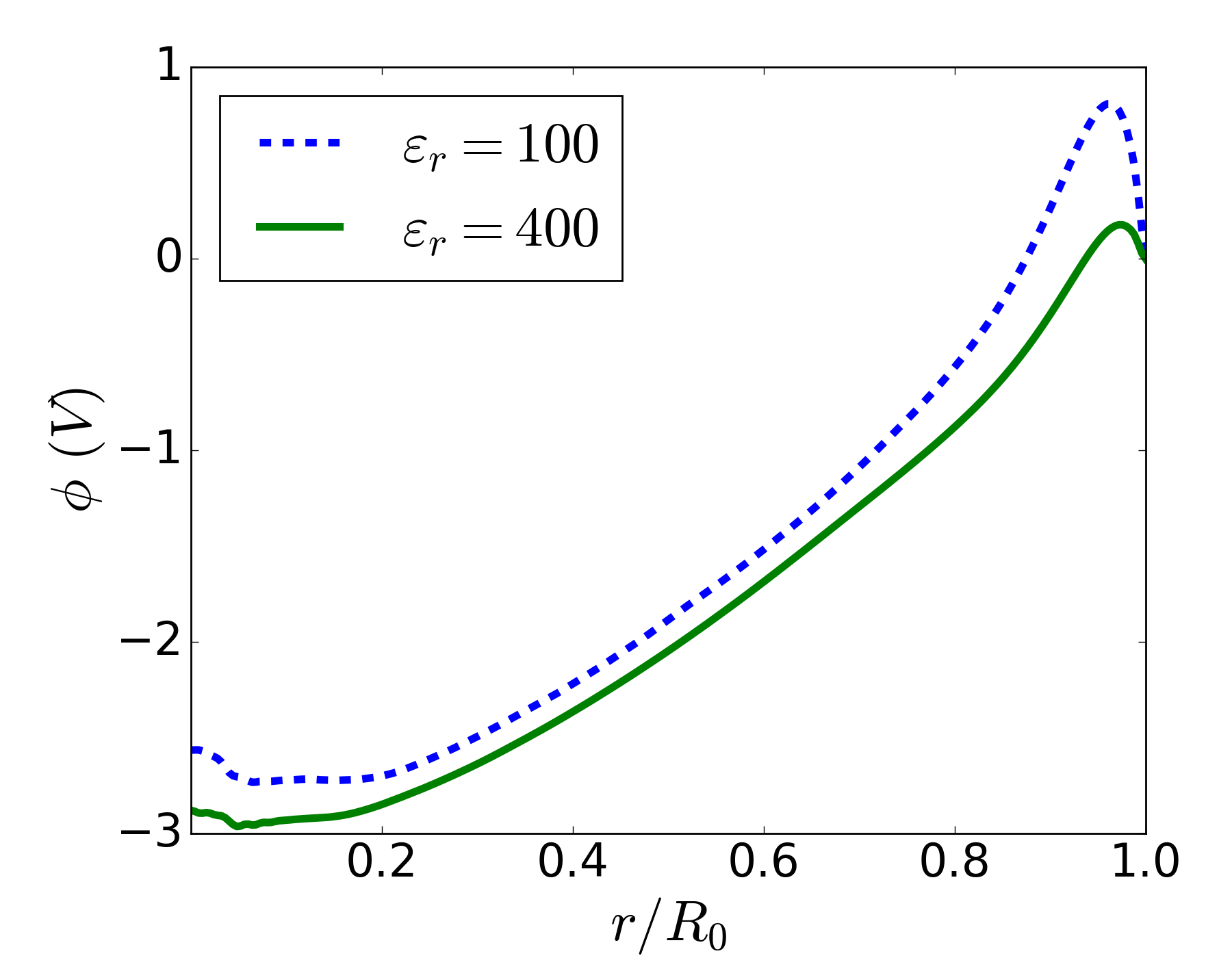}
\caption{\label{fig:avg_potential}}
\end{subfigure}

\begin{subfigure}[b]{.45\linewidth}
\includegraphics[width=\linewidth]{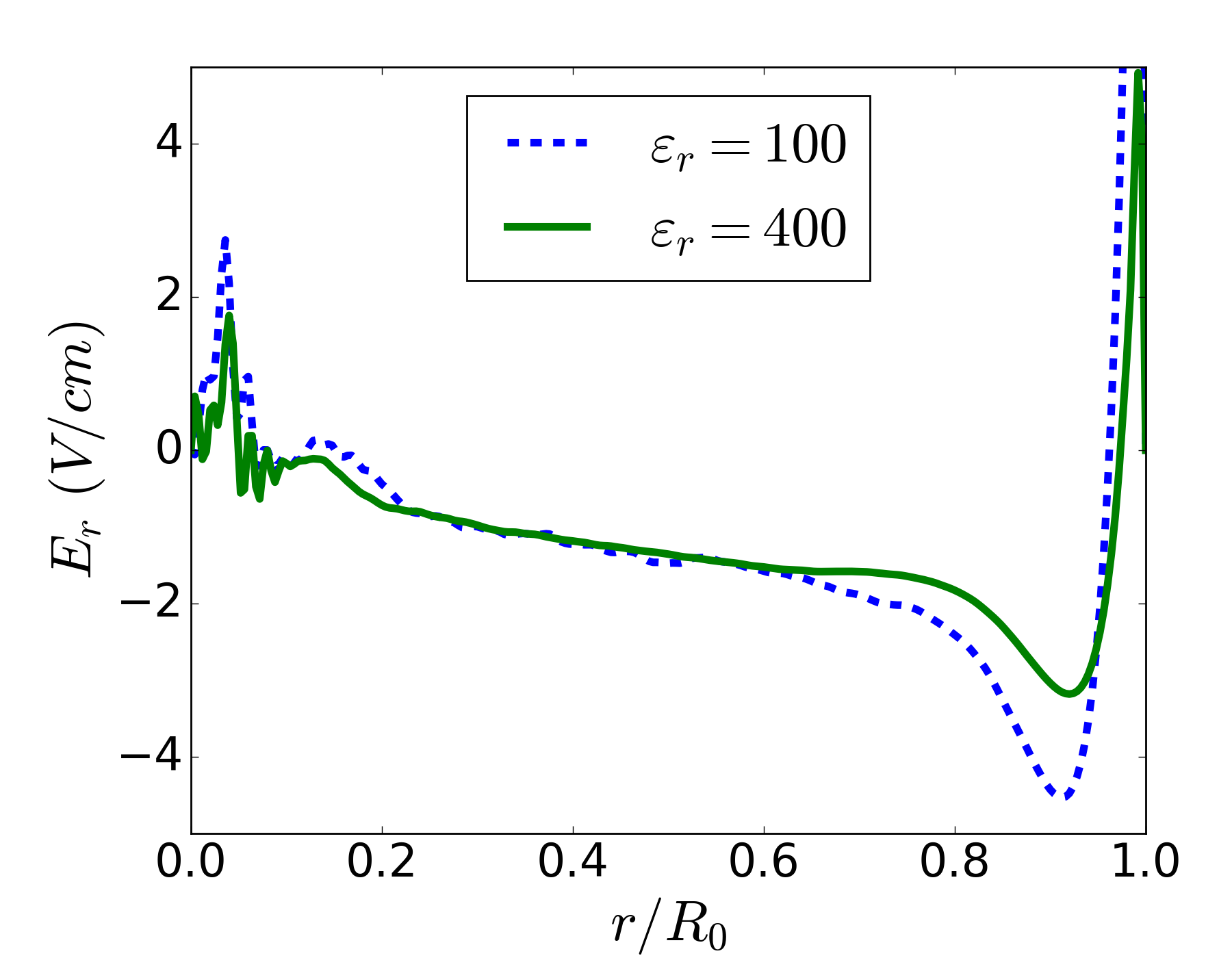}
\caption{\label{fig:avg_Efield}}
\end{subfigure}

\caption{Azimuthally and temporally averaged radial profiles of a) electric potential $\phi(r)$, b) radial electric field $E_r(r)$ for $\varepsilon_r = \{100,400\}$.}
\label{fig:avg_potential_plots}
\end{figure}

Figures \ref{fig:avg_potential} and \ref{fig:avg_Efield} show the averaged radial potential $\phi(r)$ and radial electric field $E_r(r)$ profiles. Since the unmagnetized ions are more mobile than electrons across the field lines, an ambipolar field forms in the negative radial direction. This results in a potential well at the center of the domain acting to confine ions and thereby maintain quasi-neutrality. The potential and electric field profiles are similar for different relative permittivities, with the greatest difference occurring near the sheath at the edge of the simulation $r>0.8R_0$. Since the plasma is non-neutral in this region, the sheath size and shape are strongly influenced by the relative permittivity. No physical explanation can be given for the noisy signal of average radial electric field within $r<0.1R_0$, however it is most likely either a physical or numerical artifact of particle injection.

The average electron temperature was similarly calculated and found to vary only weakly with radius, remaining near the injection temperature with an average $\left< T_e \right> = 4.7$ $eV$. Ions were found to be heated from their original temperature, with an average temperature of $\left< T_i \right> = 0.58$ $eV$. This heating is the mechanism by which ions are eventually able to escape the potential well.

To obtain estimates for the $\mathbf{E} \times \mathbf{B}$ and diamagnetic drift velocities, the radial electric field and gradient length scales are averaged away from the injection and sheath regions within $r \in [0.2 R_0, 0.8 R_0]$, giving an average azimuthal $\mathbf{E} \times \mathbf{B}$ velocity of $v_0 = \left< | E_r | \right> /B_0 = 13.1$ $km/s$ and electron diamagnetic drift velocity of $v_* = T_e/(e \left< | L_n | \right> B_0) = 64.5$ $km/s$. Within the same frame of reference as Figure \ref{fig:lin_cless}, both of these drifts occur in the anti-clockwise direction. The rotation velocity of the large scale structure is an order of magnitude smaller than the electron diamagnetic drift velocity and less than half of the $\mathbf{E} \times \mathbf{B}$ velocity. Since the large scale structure is shown to be low-frequency, long-wavelength and rotating in the $\mathbf{E} \times \mathbf{B}$ direction it exhibits all of characteristic behaviour of the plasma spoke, as observed within the Penning discharge and numerous other experiments \cite{janes1966anomalous,raitses2015effects,parker2010transition,mcdonald2011rotating}. Therefore it is proposed that the rotating structure observed within these simulations is the plasma spoke.

\subsection{Anomalous Transport Through the Rotating Spoke}

\begin{figure}[H]
\centering
\includegraphics[width=0.45\linewidth]{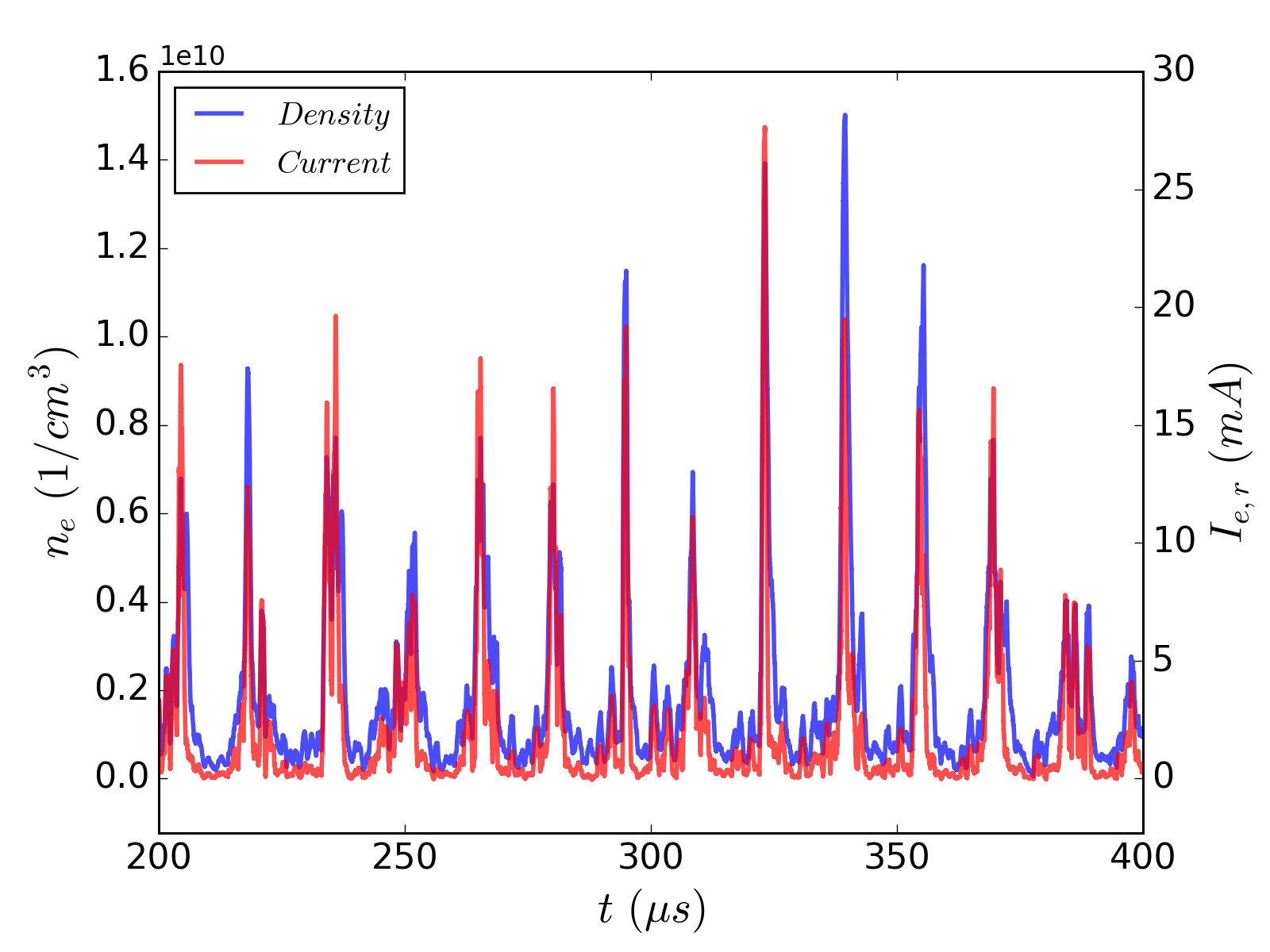}
\caption{Correlation between electron density $n_e$ and local radial electron current $I_{e,l}$ at a fixed azimuthal location on the anode.}
\label{fig:nvsi}
\end{figure}

Current and density probes were placed at various azimuthal locations near the discharge anode. Figure \ref{fig:nvsi} plots the local electron density $n_e$ at $(x_p,y_p)=(0,-0.996 R_0)$ and the magnitude of the local electron current $I_{e,l}$ passing through a horizontal chord which intersects $(x_p,y_p)$. The peaks in electron density have the same frequency to that of the spoke shown in Figure \ref{fig:lin_cless_a} and correlate with the passage of the structure. Peaks in local electron current are strongly correlated to peaks in electron density and therefore to the passage of the spoke, indicating that electron transport is enhanced by the spoke, in agreement with previous experimental observations \cite{ellison2011direct,ellison2012cross}.

The average cross-field electron conductivity $\sigma_{\perp e} (r)$ can be calculated via Ohm's law with the average electric field and gradient length scale profiles from Section \ref{first_spoke},

\begin{equation}
j_{re}(r) = \frac{I_d}{2 \pi r d} = \sigma_{\perp e} (r) \left( E_r(r) + \frac{T_e}{e L_n(r)} \right).
\label{eq:ohms}
\end{equation}

Where $j_{re}(r)$ is the cross-field electron current density and $d$ is the simulation depth (in LSP $d=1$ $cm$).

The conductivity is related to an effective turbulent collision frequency $\nu_t (r)$ via,

\begin{equation}
\sigma_{\perp e} (r) = \frac{n_e(r) e^2}{m_e \nu_t (r) (1 + \omega_c^2/\nu_t(r)^2)} \approx \frac{n_e(r) m_e \nu_t(r)}{B^2}.
\label{eq:omega}
\end{equation}

Where $\omega_c = q_e B/m_e$ is the electron cyclotron frequency and it is assumed that $\omega_c \gg \nu_t$.

\begin{figure}[H]
\centering
\includegraphics[width=0.45\linewidth]{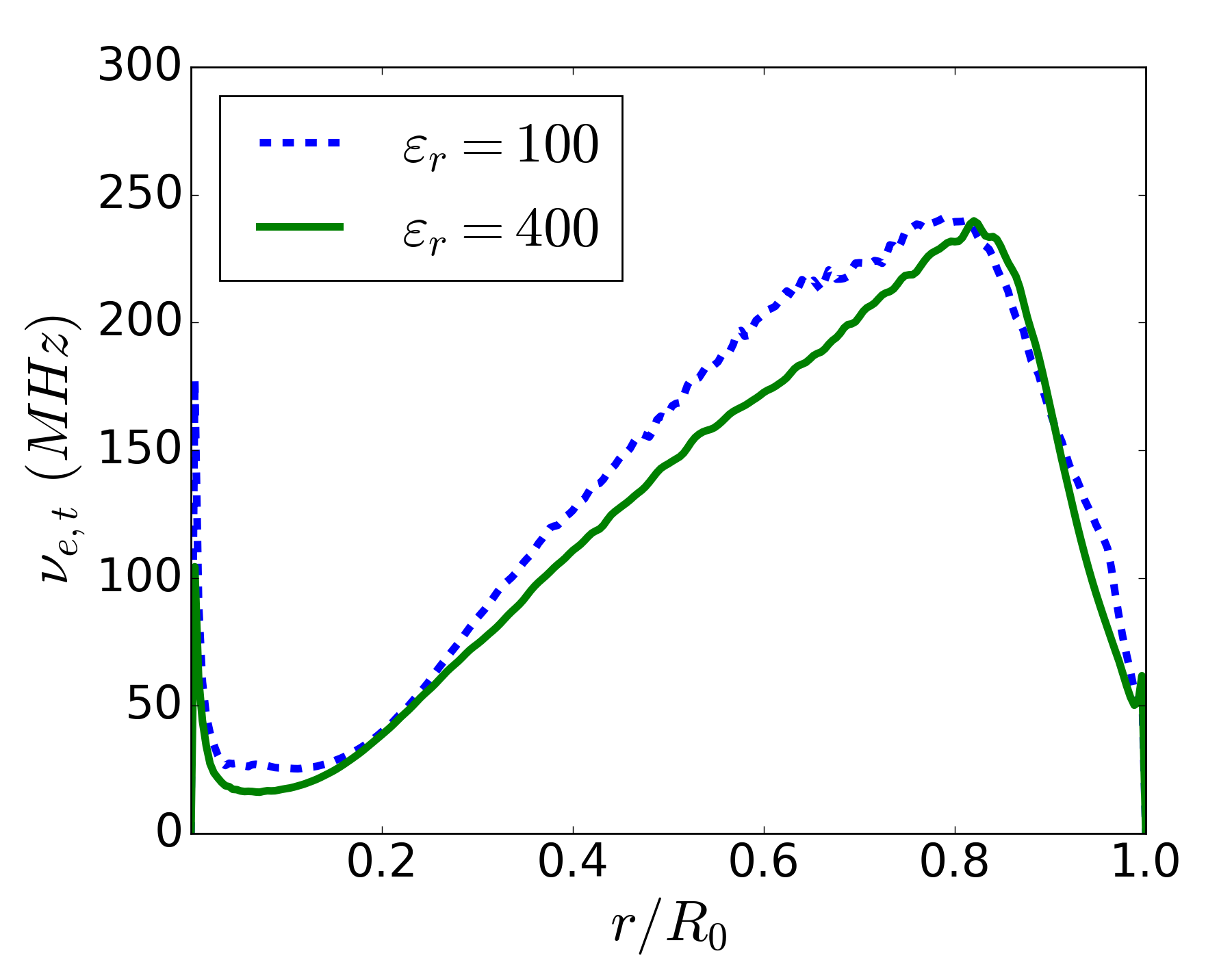}
\caption{Azimuthally and temporally averaged profiles of effective turbulent collision frequency $\nu_t(r)$ for $\varepsilon_r = \{100,400\}$.}
\label{fig:avg_cfreq}
\end{figure}

Figure \ref{fig:avg_cfreq} plots the average turbulent collision frequency as a function of radius. Values are similar for both the $\varepsilon_r=100$ and $\varepsilon_r=400$ cases. Within $r \in [0.2 R_0, 0.8 R_0]$ the mean turbulent collision frequency for the $\varepsilon_r=400$ case is $\left< \nu_t \right> = 139.5$ $MHz$, giving an effective Hall parameter $\beta_{eff} = \omega_c/\nu_t=12.6$. This value is comparable to those measured by Bohm within numerous experiments exhibiting anomalous transport \cite{bohm1949characteristics}. The decay of the collision frequency for $r > 0.8R_0$ is most likely due to non-neutral behaviour within the sheath.

\subsection{Collisional Simulations of the Rotating Spoke}
\label{spoke_collision}

To better replicate conditions of the experiment, the collisionless simulation from Section \ref{first_spoke} is modified to allow for ionization via MCC processes. The simulation domain initially consists of a fixed uniform background of neutral particles with pressure $P_n$ and temperature $T_n$ (see Table \ref{tab:sim_params}). A uniform axial beam of electrons is injected with energy $V_b$ and ions are formed via collisional processes between electrons and neutrals, note that neutrals are not depleted during this process.

The plasma is initially non-neutral, resulting in highly excited electrons which rapidly ionize the background gas and render the system quasi-neutral. A large scale, low-frequency single mode develops after $20$ $\mu s$, rotating in the $\mathbf{E} \times \mathbf{B}$ direction. Figure \ref{fig:lin_cless} shows four contour plots of electron density, commencing at $61.4$ $\mu s$ and incremented by a $\pi/4$ phase shift of the rotating structure. This structure exhibits all of the same physical properties as in the collisionless case and is similarly proposed to be the plasma spoke.

\begin{figure}[H]
\centering
\includegraphics[width=1.0\linewidth]{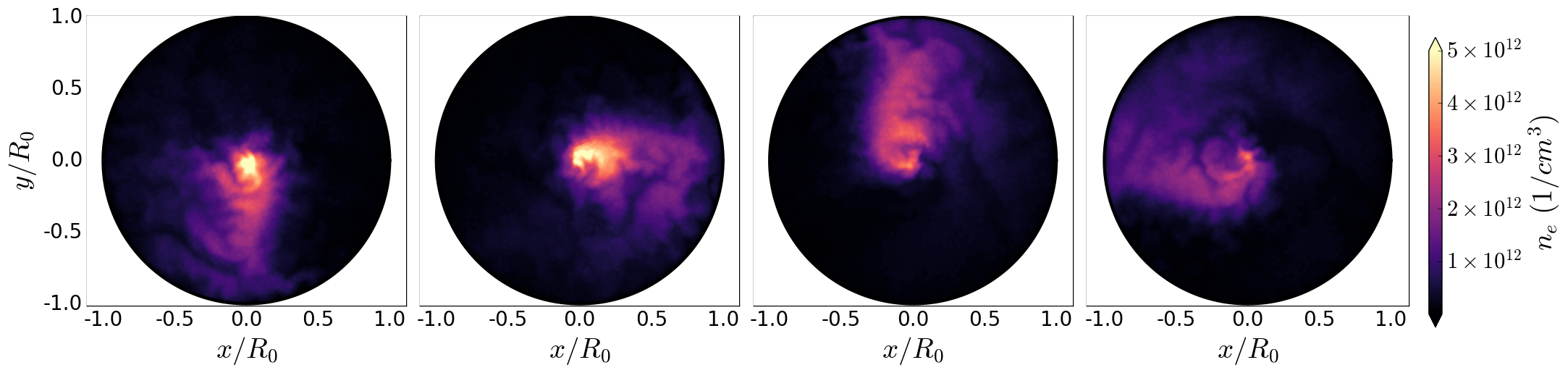}
\caption{\label{fig:nlin_ionize}Electron density contours of the collisional Penning discharge at simulation times, from left to right; $61.4$ $\mu s$, $65.8$ $\mu s$, $70.7$ $\mu s$ and $74.2$ $\mu s$.}
\end{figure}

The spoke frequency is computed as $f_s = 62.4$ $kHz$ with a rotation velocity at $r=R_0/2$ of $v_r=4.90$ $km/s$, both within $6 \%$ of the values obtained for the collisionless case. The structure of the density contours reveal similar behaviour to that of the collisionless case, although the spoke structure extends further towards the anode and appears to have a larger azimuthal extent. This most likely indicates that ionization is occurring not only at the center of the trap, but also within the spoke itself, enhancing plasma density within this region and broadening the spoke.

\begin{figure}[H]
\centering
\begin{subfigure}[b]{.45\linewidth}
\includegraphics[width=\linewidth]{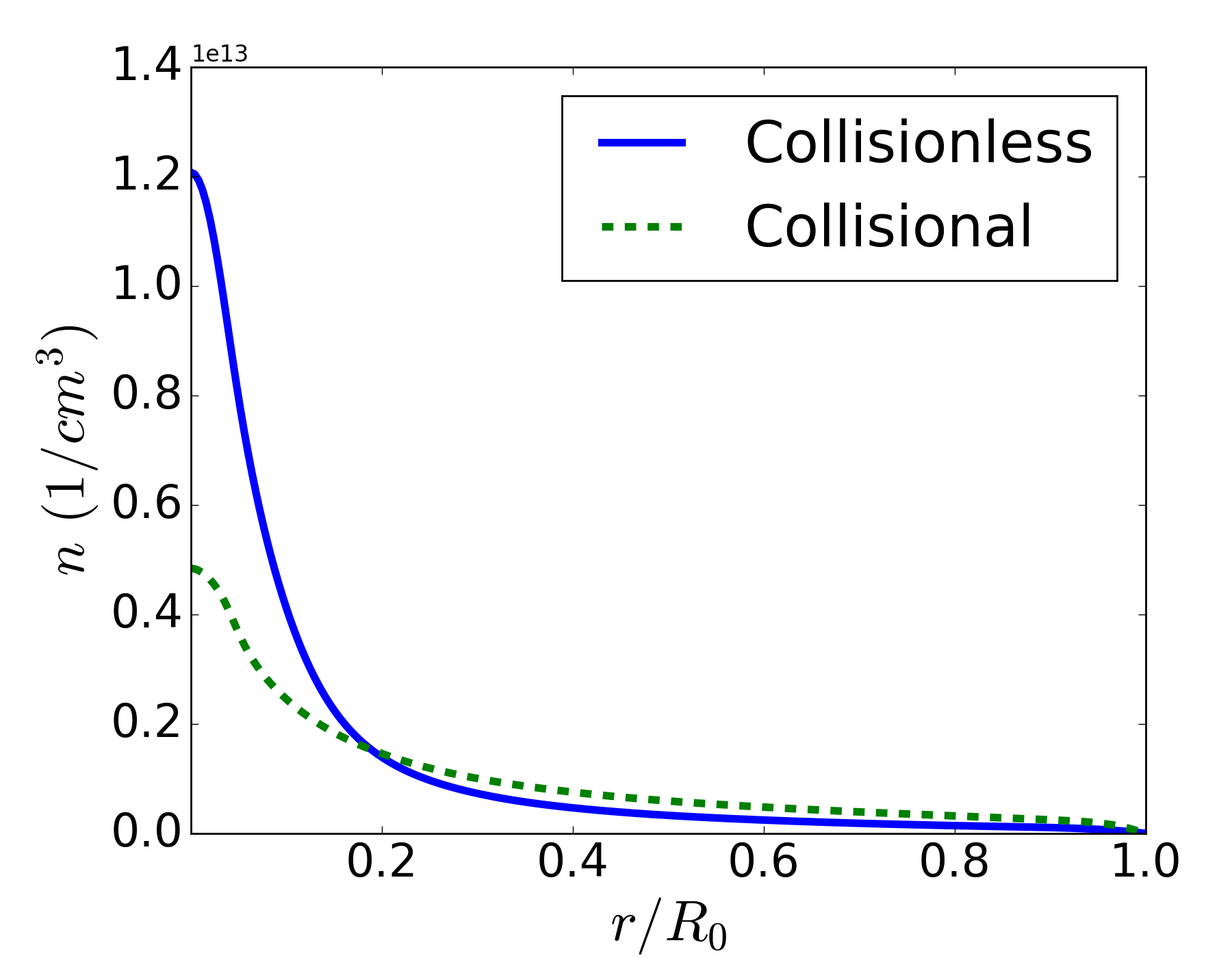}
\caption{\label{fig:avg_density_ionize}}
\end{subfigure}

\begin{subfigure}[b]{.45\linewidth}
\includegraphics[width=\linewidth]{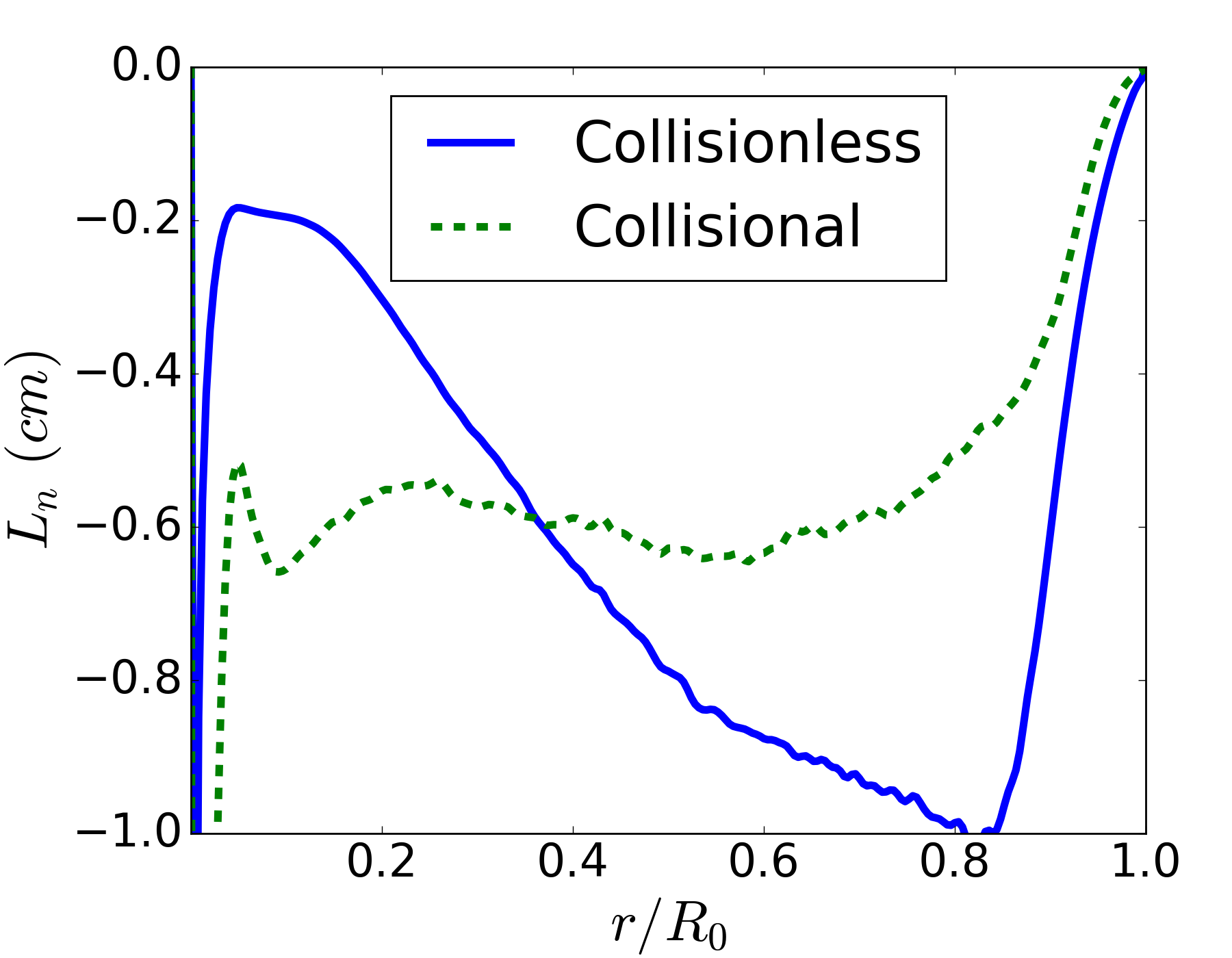}
\caption{\label{fig:avg_gls_ionize}}
\end{subfigure}

\caption{Azimuthally and temporally averaged radial profiles of a) electron density $n_e(r)$, b) gradient-length-scale $L_n(r)=n_e(r)/(dn_e(r)/dr)$ for collisional and collisionless cases.}
\label{fig:avg_density_plots_ionize}
\end{figure}

\begin{figure}[H]
\centering
\begin{subfigure}[b]{.45\linewidth}
\includegraphics[width=\linewidth]{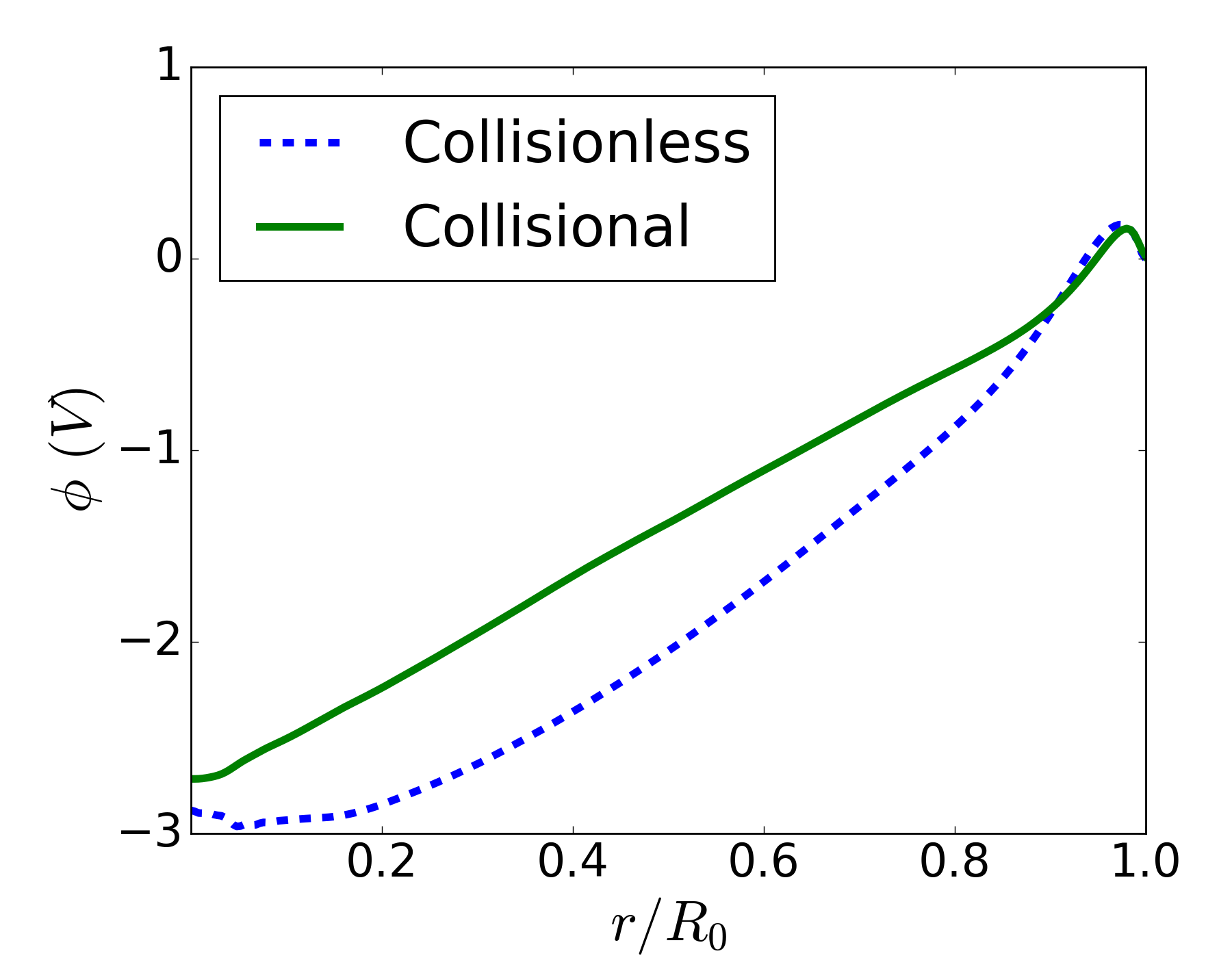}
\caption{\label{fig:avg_potential_ionize}}
\end{subfigure}

\begin{subfigure}[b]{.45\linewidth}
\includegraphics[width=\linewidth]{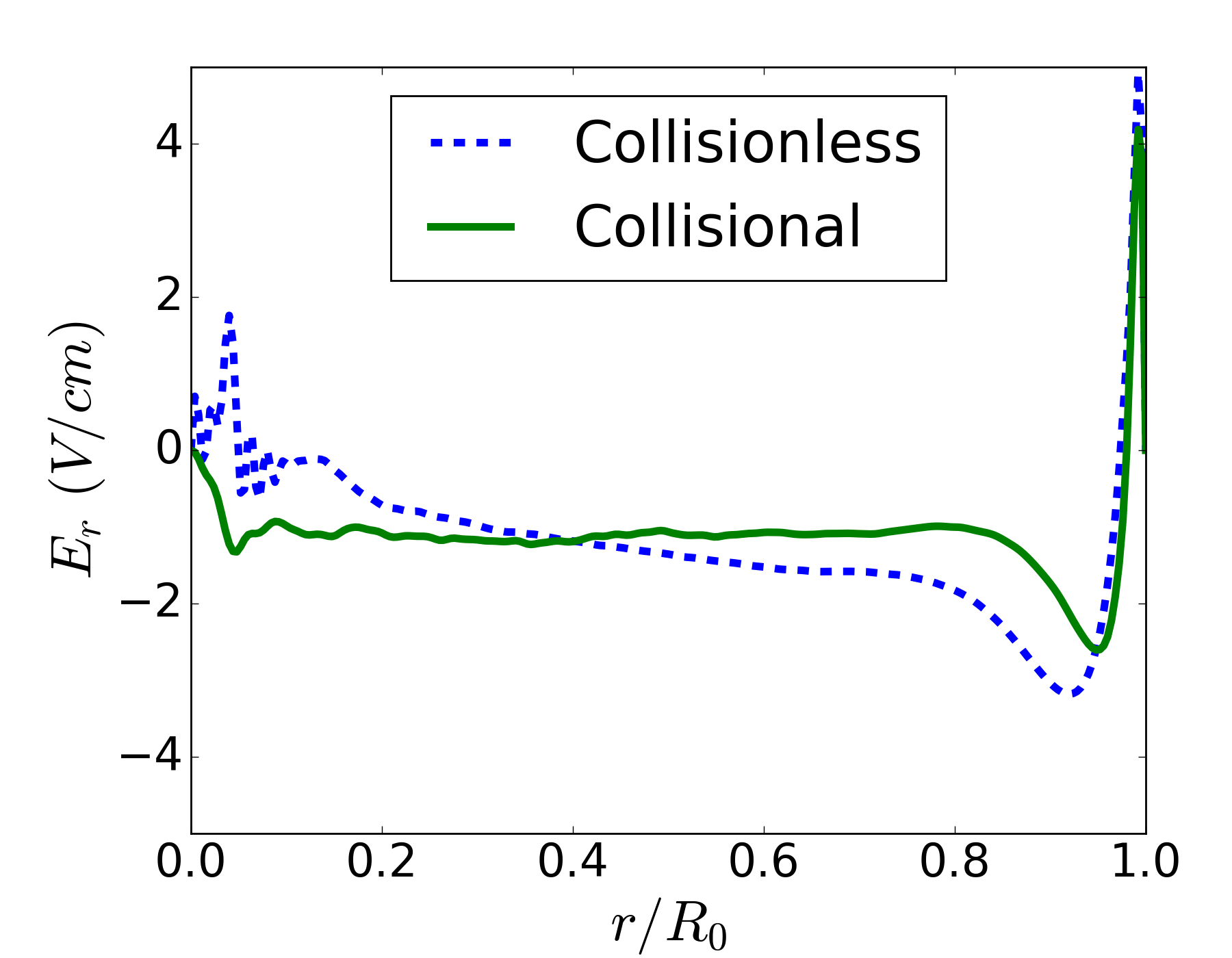}
\caption{\label{fig:avg_Efield_ionize}}
\end{subfigure}

\caption{Azimuthally and temporally averaged radial profiles of a) electric potential $\phi(r)$, b) radial electric field $E_r(r)$ for collisional and collisionless cases.}
\label{fig:avg_potential_plots_ionize}
\end{figure}

The average radial density and average gradient length scale profiles (see Figure \ref{fig:avg_density_plots_ionize}) and average electric potential and average electric field profiles (see Figure \ref{fig:avg_potential_plots_ionize}) are computed and compared with the collisionless case. The density profile and density gradients for the collisional case are shallower than the collisionless case. This also indicates that ionization is likely occurring outside of the injection region, since a more diffuse source of ions can sustain a flatter electron density profile. The electric potential and electric field profiles are similar between both cases, with the collisional case exhibiting a more linear potential and therefore flatter electric field profile.

The profiles in electric field and gradient length scale are used to compute the average effective electron collision frequency $\nu_{eff}=107.2$ $MHz$ and effective Hall parameter $\beta_{eff}=16.4$, similar in order of magnitude to the collisionless case. The electron-neutral collision frequency for Xenon with $T_e \approx 5$ $eV$ is $\nu_{en}=12.2$ $MHz$. This gives a classical Hall parameter of $\beta_{c}=144.3$ indicating that the transport is highly anomalous and dominated by turbulent fluctuations.

Despite these differences, the spoke frequency and structure as well as the average discharge profiles are remarkably similar to the collisionless case. This indicates that the same fundamental mechanism is likely responsible for the formation of the spoke. Since it was shown in Section \ref{first_spoke} that ionization is not necessary for spoke formation, it is unlikely that the spoke is caused by an ionization wave. Furthermore for the reduced mass Xenon, the CIV \cite{alfven1967origin} $v_{civ} = 17.1$ $km/s$, significantly larger than the spoke rotation velocity.

This leaves the CSHI as a likely candidate for the formation of the spoke within these simulations. In both the collisional and collisionless cases $E_r(r) / L_n(r) > 0 \; \forall \; r$ such that the instability criterion for the CSHI is satisfied. Keeping in mind the limitations of these simulations it is therefore possible that the rotating spoke observed within the Penning discharge is the result of the CSHI, rather than an ionization wave.

\subsection{Spoke Frequency Scaling with Discharge Parameters}
\label{parameter_scan_discharge_current}

Considering the relative magnitudes of the average diamagnetic drift velocity $v_* = 64.5$ $km/s$, $\mathbf{E} \times \mathbf{B}$ velocity $v_0 = 13.1$ $km/s$ and ion-sound speed $v_s = 10.6 $ $km/s$, linear CSHI theory suggests the following scaling for the spoke angular frequency \cite{smolyakov2016fluid},

\begin{equation}
\omega_{s,th} = \sqrt{\frac{v_s^2 v_0}{v_*} k^2} = \sqrt{\frac{e E_r L_n}{m_i} k^2}.
\end{equation}

Assuming a single azimuthal mode propagating at $r=R_0/2$, we have $k = k_{\theta} = 2/R_0$, and a theoretical estimate for the spoke frequency,

\begin{equation}
f_{s,th} = \frac{1}{\pi R_0} \sqrt{\frac{e E_r L_n}{m_i}}.
\label{eq:fapprox}
\end{equation}

The validity of this scaling is tested by modifying the discharge parameters of the collisionless simulation introduced in Section \ref{first_spoke}. For each simulation, the radial electric field multiplied by the gradient length scale $\left< | E_r L_n | \right>$ is averaged for $r \in [0.2 R_0, 0.8 R_0]$. The estimate for spoke frequency $f_{s,th}$ is plotted with the measured spoke frequency $f_s$ against different discharge parameters.

\begin{figure}[H]
\centering
\includegraphics[width=0.45\linewidth]{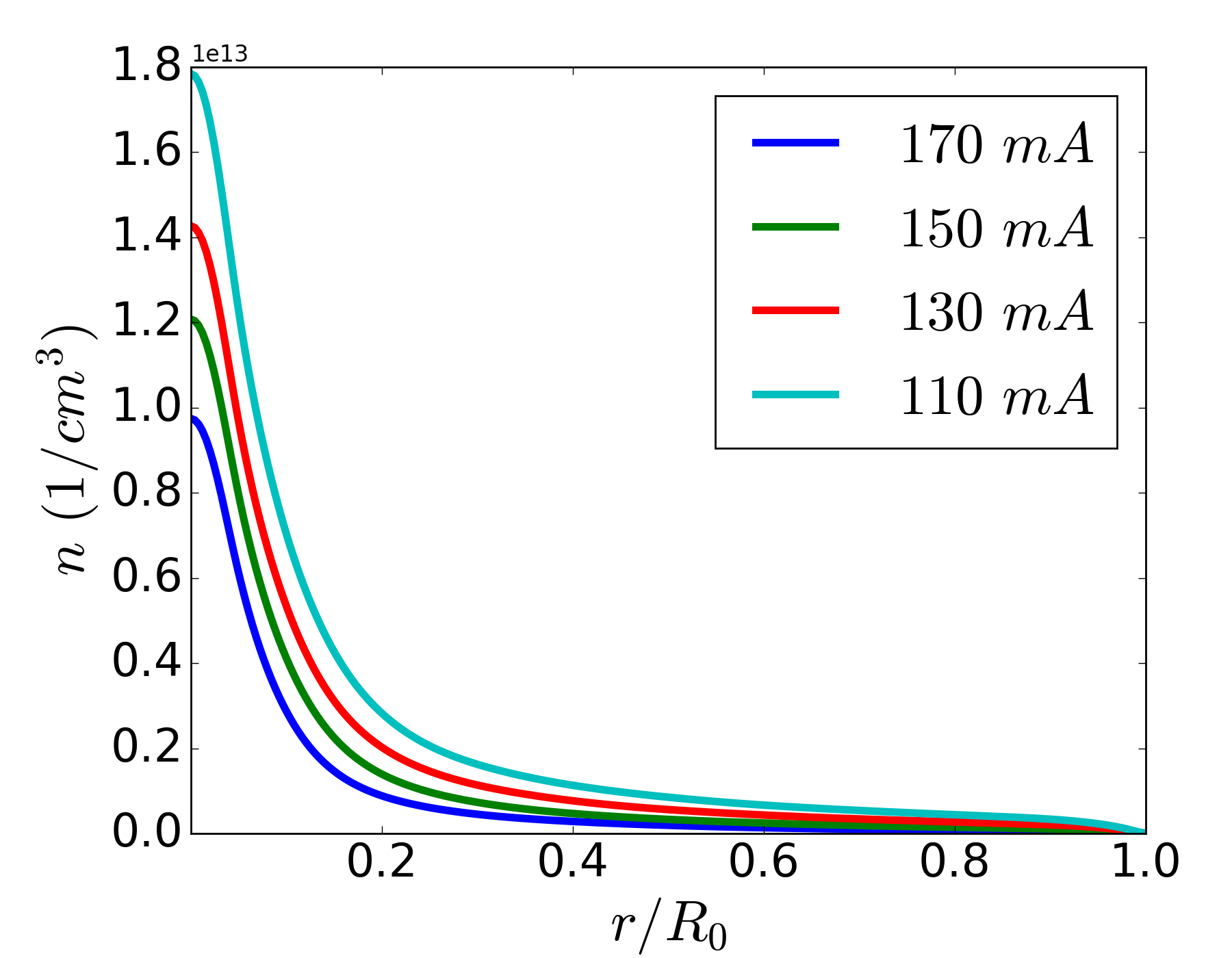}
\caption{Temporally and azimuthally averaged radial density profile $n_e(r)$ for various discharge currents.}
\label{fig:cscan_dprofile}
\end{figure}

\begin{figure}[H]
\centering
\includegraphics[width=0.45\linewidth]{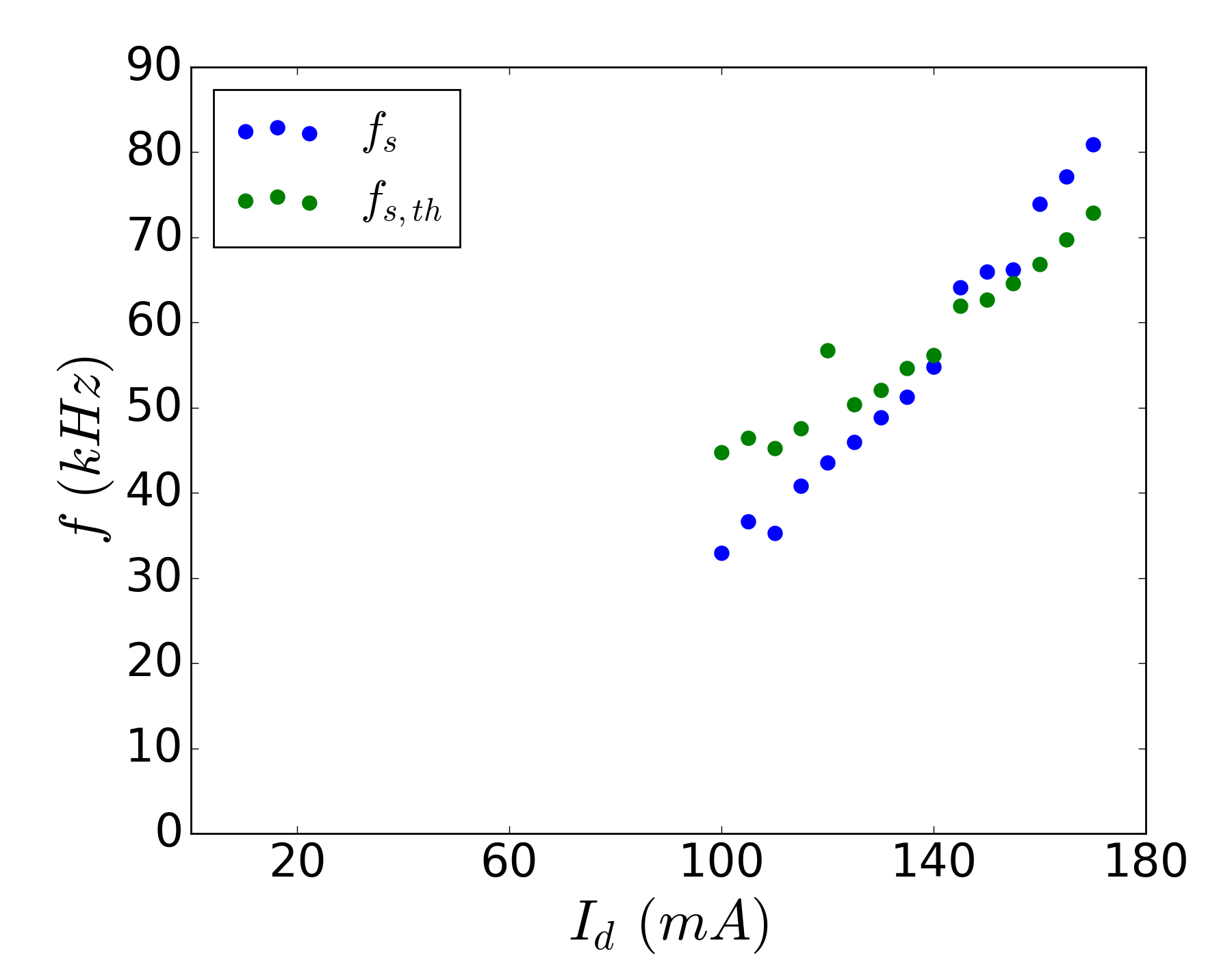}
\caption{Spoke rotation frequency $f_s$ and predicted rotation frequency $f_{s,th}$ as a function of discharge current $I_d$.}
\label{fig:cscan_f}
\end{figure}

The discharge current $I_d$ is varied by modifying the ion injection current $I_{i}$. Figure \ref{fig:cscan_dprofile} shows how the average radial density profile changes with discharge current. Increasing the discharge current results in relatively fewer background ions available to neutralize injected electrons and therefore a shallower density profile. The density at the center of the trap $n_e(0)$ correlates linearly with discharge current. Figure \ref{fig:cscan_f} then demonstrates a correlation between spoke frequency and discharge current.

\begin{figure}[H]
\centering
\includegraphics[width=0.45\linewidth]{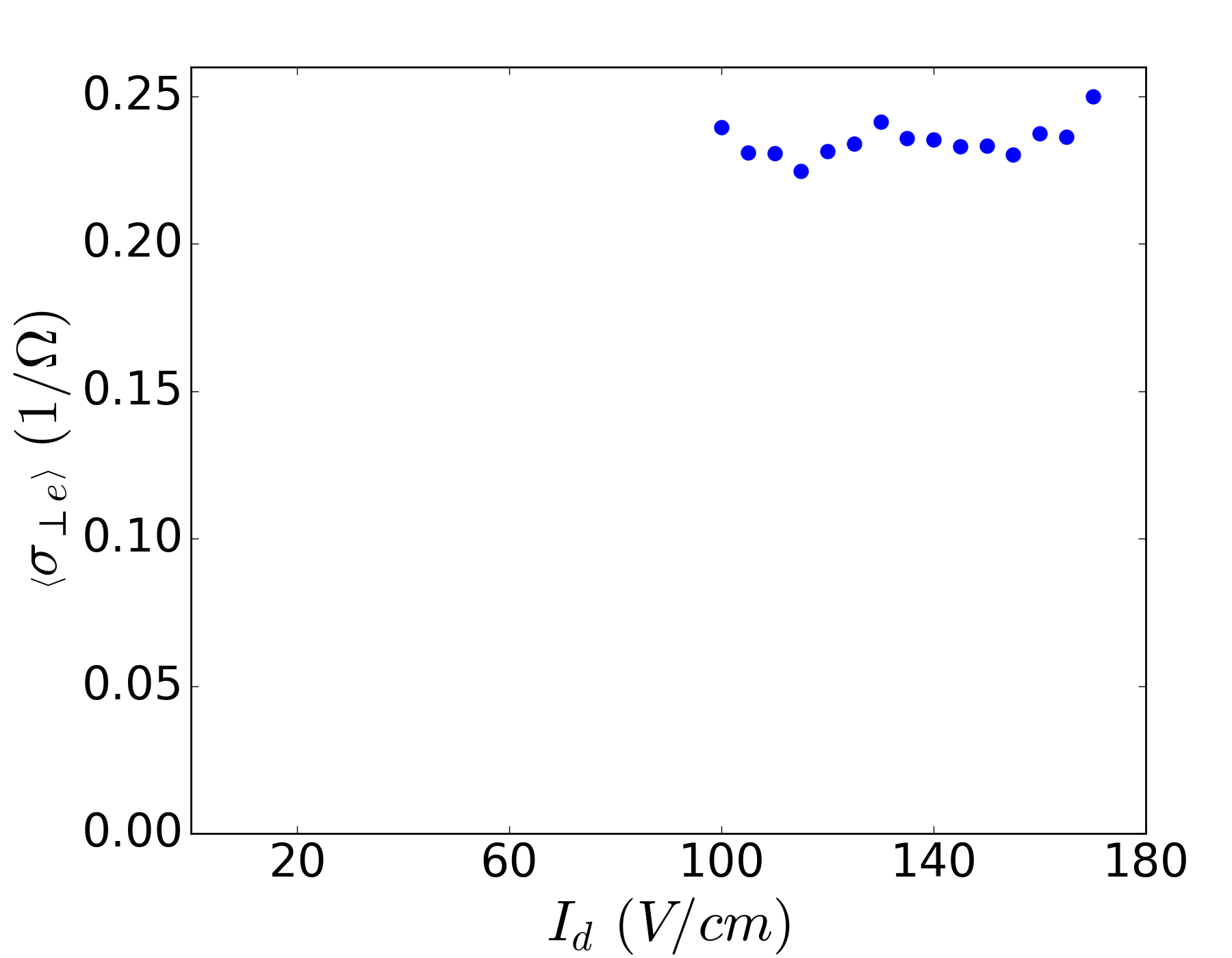}
\caption{Average radial electron cross-field conductivity $\left< \sigma_{\perp e} \right>$ as a function of discharge current $I_d$.}
\label{fig:cscan_sigma}
\end{figure}

Additionally, the average electron cross-field conductivity is computed at $r=R_0/2$ via Ohm's law (Equation \ref{eq:ohms}) and plotted with respect to the discharge current in Figure \ref{fig:cscan_sigma}.

\begin{figure}[H]
\centering
\includegraphics[width=0.45\linewidth]{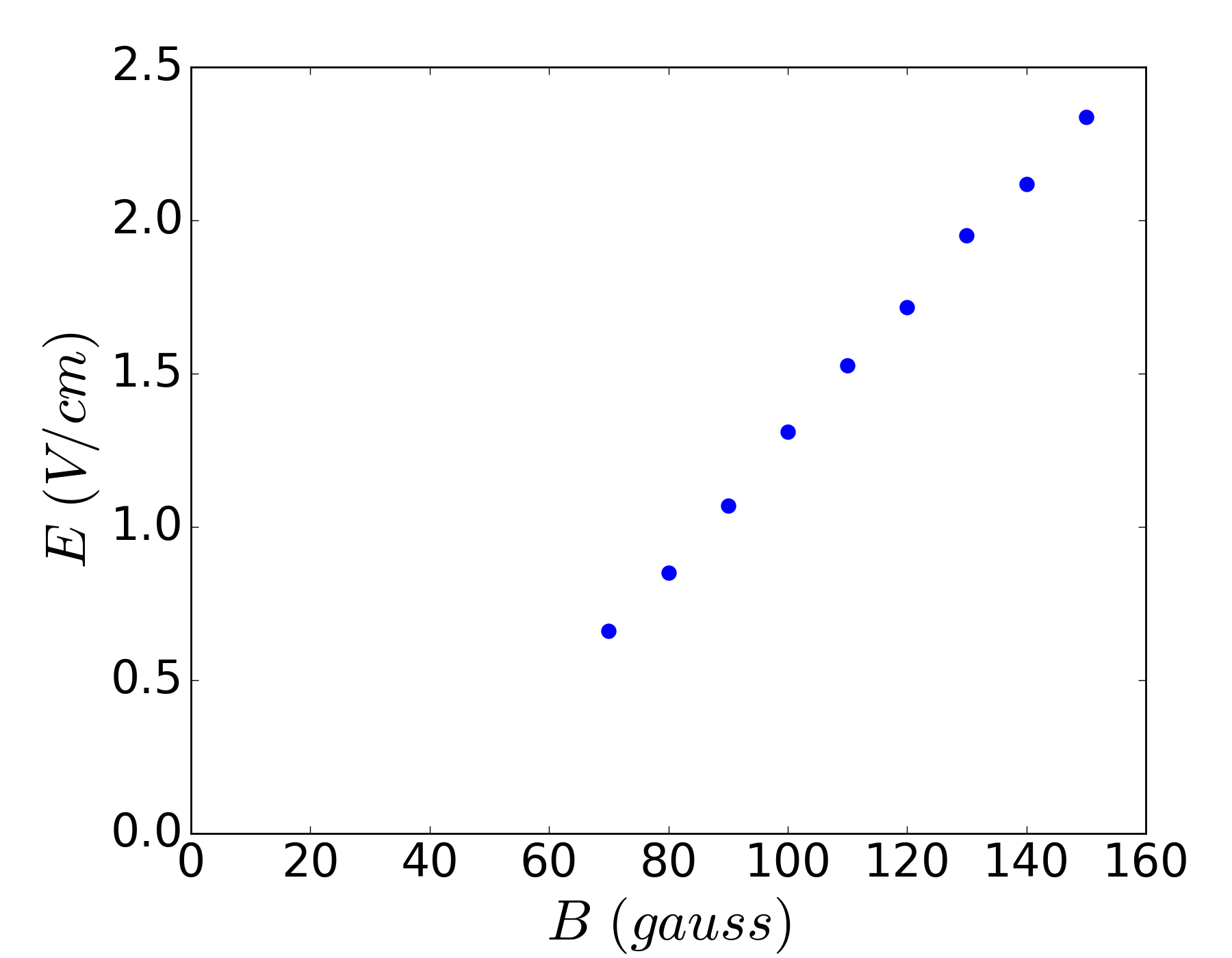}
\caption{Average radial electric field $\left< | E_r | \right>$ as a function of applied magnetic field strength $B$.}
\label{fig:bscan_efield}
\end{figure}

\begin{figure}[H]
\centering
\includegraphics[width=0.45\linewidth]{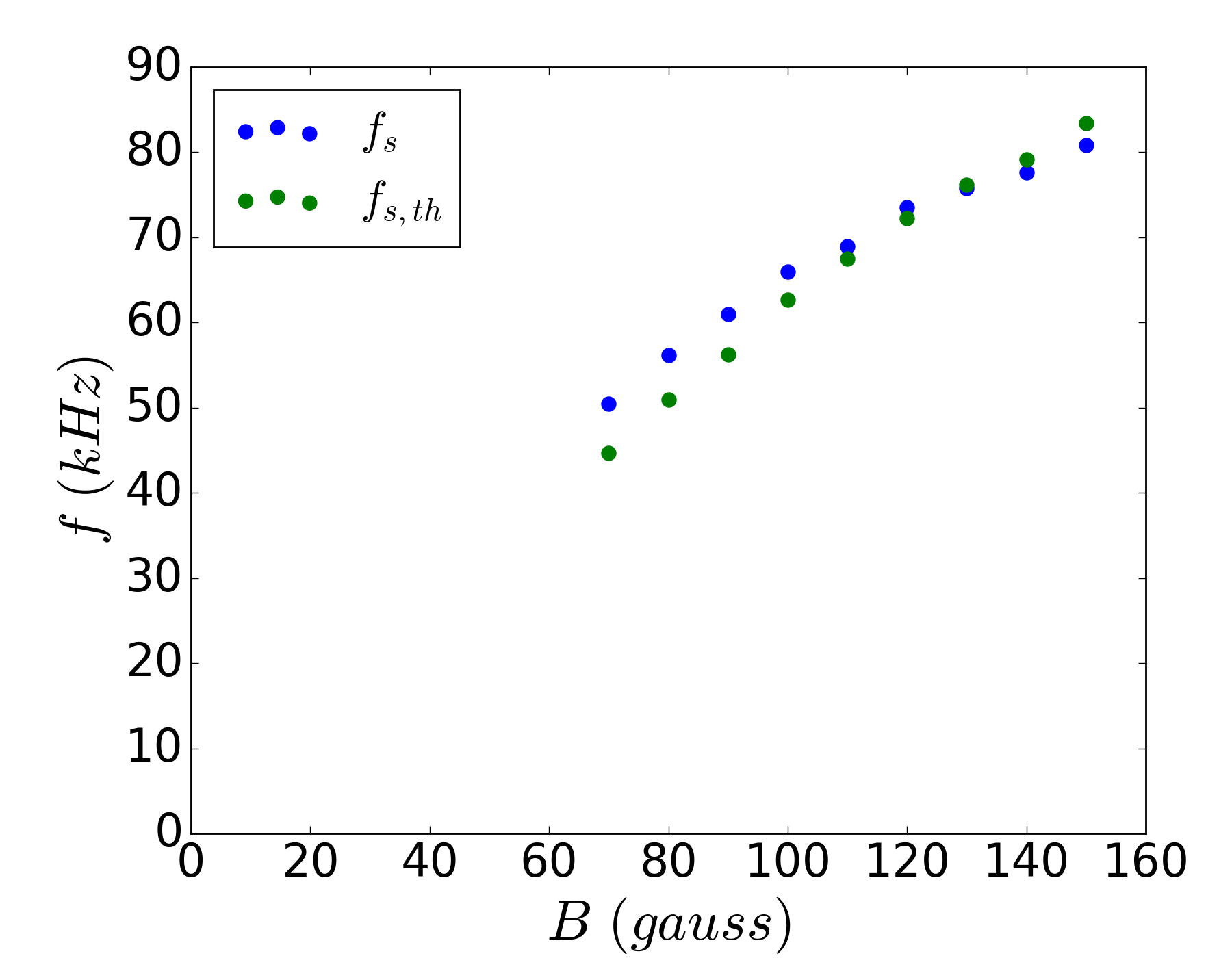}
\caption{Spoke rotation frequency $f_s$ and predicted rotation frequency $f_{s,th}$ as a function of applied magnetic field strength $B$.}
\label{fig:bscan}
\end{figure}

Increasing the applied magnetic field strength reduces electron mobility and therefore results in an enhanced ambipolar electric field. Figure \ref{fig:bscan_efield} shows a near linear correlation between the radial electric field and applied magnetic field strength. Therefore, as per Equation \ref{eq:fapprox}, the spoke frequency should scale as $f_s \sim \sqrt{B}$. This is demonstrated in Figure \ref{fig:bscan} and consistent with experimental observations \cite{raitses2015effects}.

\begin{figure}[H]
\centering
\includegraphics[width=0.45\linewidth]{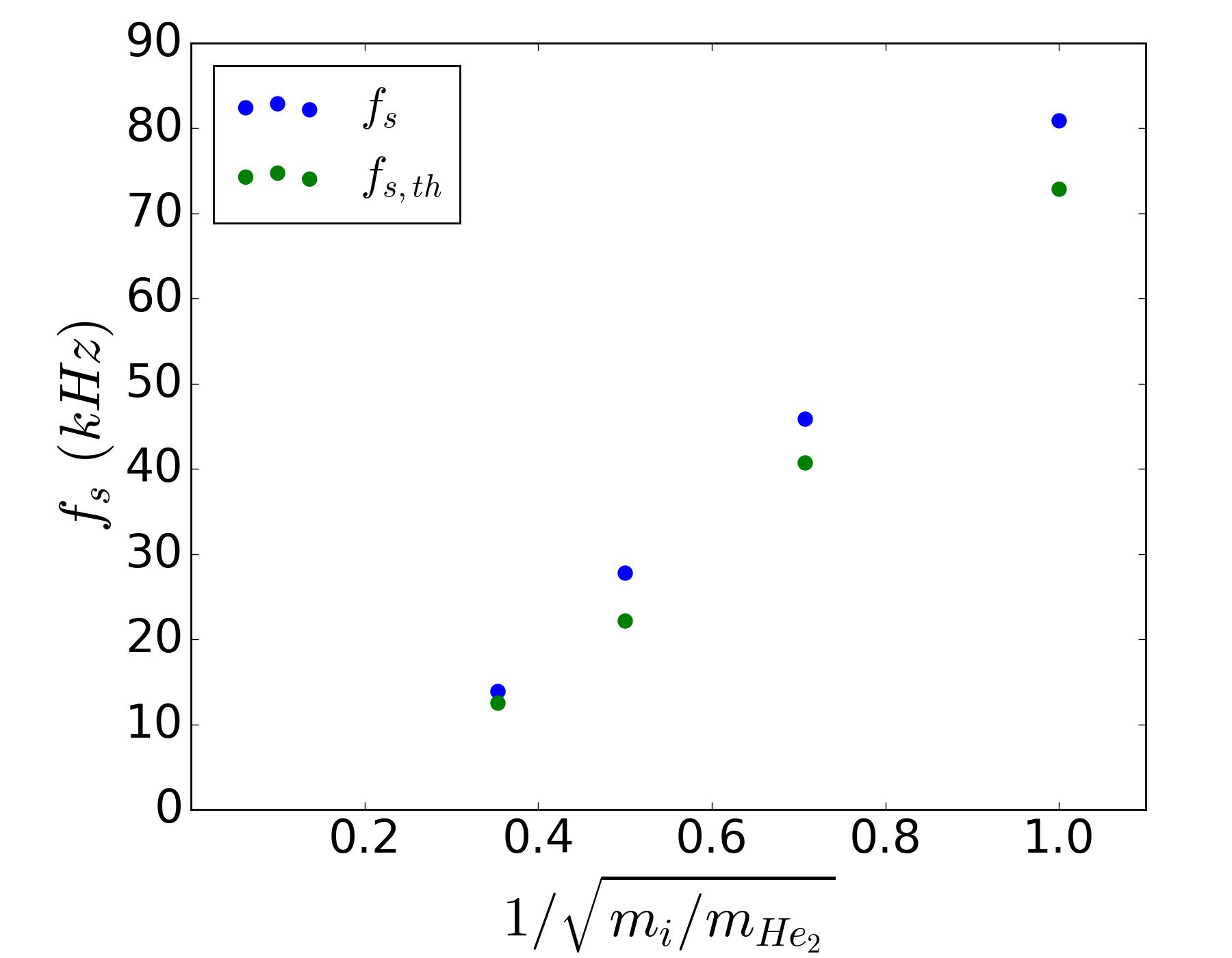}
\caption{Spoke rotation frequency $f_s$ and predicted rotation frequency $f_{s,th}$ as a function of the inverse square root of ion mass $m_i$.}
\label{fig:mscan}
\end{figure}

Ion-mass $m_i$ is varied and Figure \ref{fig:mscan} shows a near linear correlation between spoke frequency and $1/\sqrt{m_i}$ (normalized to the ion mass of Helium-4), consistent with experimental observation \cite{raitses2015effects}. This also demonstrates that the spoke rotation velocity scales in an identical way to the ion-acoustic velocity with respect to ion-mass.

For each of these parameter scans the approximate theoretical estimate for the spoke frequency shows good agreement in both magnitude and scaling to the measured numerical spoke frequency, providing strong evidence for this scaling relationship and that the CSHI is the driving instability for the rotating spoke. It should be taken into consideration that the observed structure is clearly highly non-linear and turbulent, making it surprising that a simple estimate based on linear theory provides such an accurate fit.

\section{Conclusion}

A highly non-linear turbulent structure rotating in the azimuthal $\mathbf{E} \times \mathbf{B}$ direction is observed to form within full-size two-dimensional kinetic simulations of a Penning discharge. This structure exhibits characteristic behaviour very similar to that of the rotating spoke observed within experiments. Electron cross-field transport within the discharge is highly anomalous, with a majority of the current being channeled through the spoke structure. The fact that the spoke forms without ionization and the correlation of the resulting spoke frequency with $\sqrt{e E_r L_n / m_i}$ suggests that the collisionless Simon-Hoh instability is responsible for its formation.

\section*{Acknowledgements}

Physics work was supported by the Air Force Office of Scientific Research.

Code development was supported by the Princeton University Program in Plasma Science and Technology.

This research used resources of the National Energy Research Scientific Computing Center, a DOE Office of Science User Facility supported by the Office of Science of the U.S. Department of Energy under Contract No. DE-AC02-05CH11231.

This research used resources of the Perseus cluster at the TIGRESS high performance computer center at Princeton University, which is jointly supported by the Princeton Institute for Computational Science and Engineering and the Princeton University Office of Information Technology's High Performance Research Computing Center.

\newpage

\bibliographystyle{ieeetr}
\bibliography{sample}

\end{document}